\documentclass[pre,aps,10pt,amsmath,amsfonts,amssymb,twocolumn]{revtex4-2}
\usepackage{graphicx,color,mathtools}
\usepackage{CJK}
\usepackage{orcidlink}
\usepackage[normalem]{ulem}
\newcommand{\C}{\mathcal{C}}
\renewcommand{\P}{\mathbb{P}}
\newcommand{\fA}{\mathcal{A}}
\newcommand{\fB}{\mathcal{B}}
\newcommand{\M}{\mathfrak{A}}
\begin{document}
\begin{CJK*}{UTF8}{mj}
\title{Effective Hardcore Exclusion Without Exclusion: Two-Species Pair Annihilation Revisited}

\author{Su-Chan Park (박수찬)\,\orcidlink{0000-0002-0890-8879}}
\affiliation{Department of Physics, The Catholic University of Korea, Bucheon 14662, Republic of Korea}
\author{Foster Thompson\,\orcidlink{0000-0003-1844-5852}}
\affiliation{Institut f\"ur Theoretische Physik, Universit\"at zu K\"oln, 50937 K\"oln, Germany}

\date{\today}
\begin{abstract}
Reaction-diffusion systems with two species that mutually annihilate through $A+B\to\emptyset$ reactions display a slow decay of the total density $\rho$ which follows an anomalous power law in low dimensions. For hardcore particles in one dimension, this decay follows $\rho \sim t^{-1/4}$ under symmetric diffusion and $\rho \sim t^{-1/3}$ under asymmetric diffusion without relative bias between the two species. The $t^{-1/3}$ behavior, in particular, has so far been observed exclusively in systems with hardcore exclusion. Here we construct a model of particles without hardcore exclusion that reproduces this same $t^{-1/3}$ scaling. In our model, both species undergo asymmetric diffusion with a rate that depends nonlinearly on the local density, a form of transport that, in the absence of the other species, is superdiffusive and falls into the Kardar-Parisi-Zhang universality class. The scaling behavior of $t^{-1/3}$ occurs when at least one of the two species undergoes asymmetric diffusion induced by microscopic processes involving pairs of particles, while asymmetric processes involving groups of three particles lead to $t^{-1/4}$ scaling associated with symmetric diffusion.  This demonstrates that the pair annihilation density decay exponent is not determined exclusively by the transport properties of the isolated particle species.

\end{abstract}
\maketitle
\end{CJK*}
\section{\label{Sec:intro}Introduction}
Certain reaction-diffusion systems with hardcore exclusion have often been reported to exhibit scaling behavior that seems to depart from their ``bosonic'' counterparts (in which occupation numbers of particles at each lattice site are unbounded), even in the low-density regime where encounters between particles at the same site are rare~\cite{Janowsky1995,Janowsky1995b,Is1995,Kwon2000}.
One prominent example is two-species pair-annihilation dynamics, conventionally denoted by $A+B\rightarrow\emptyset$, of diffusing particles in one dimension.

If the initial densities of species $A$ and species $B$ are equal, the long-time behavior of the total density is known to follow a power law $t^{-\delta}$, with the density-decay exponent $\delta$ related to the dynamic exponent $z$ via $\delta = 1/(2z)$ in one dimension. It is well established that $z=2$ (i.e., $\delta=\frac14$) in the symmetric-diffusion case, and $z=1$ (i.e., $\delta=\frac12$) when there is a relative bias between the two species~\cite{TW1983,BL1991,KR1984,Lee1995}. In the presence of hardcore exclusion among particles of the same species, however, the density-decay exponent is argued to instead be $\delta=\frac13$ when both species undergo biased diffusion but \emph{without} a relative bias between them~\cite{Janowsky1995,Janowsky1995b,Is1995}. The origin of $t^{-1/3}$ decay has been attributed to the fact that hardcore particles subject to biased diffusion undergo superdiffusive transport with the Kardar-Parisi-Zhang (KPZ) dynamic exponent $z=\frac32$, which is argued to persist within the segregated single-species regions even at low density. In this sense, the $t^{-1/3}$ density decay has come to be regarded as a distinctive feature of hardcore systems, one that might not be realizable within bosonic models.

Intrigued by this seeming discrepancy, a number of attempts were made around the turn of the 21st century to formulate a field theory that properly incorporates the effects of exclusion, using either a complex-number (bosonic) field~\cite{Park2000,vW2001,PP2005PRE} or a Grassmann-number field~\cite{Bares1999,Brunel2000}. In the former approach, particle exclusion enters only as a particular structure within an otherwise ordinary bosonic field theory, rather than requiring a fundamentally different field content. This, in fact, suggests that scaling behavior previously thought to arise specifically from hardcore dynamics should also be accessible from bosonic \emph{microscopic} models. Motivated by this reasoning, we construct a microscopic model of bosonic particles that reproduces the $t^{-1/3}$ scaling behavior previously reported only for hardcore systems. In our bosonic model, we find that a biased nonlinear hopping process involving two particles reproduces KPZ superdiffusion, and that, when paired with the $A+B\to\emptyset$ reaction rule, it likewise reproduces the $t^{-1/3}$ density decay.

Beyond reproducing this particular result, formulating the problem in terms of bosonic particles is advantageous in its own right. One is free to design a much richer variety of interactions than the hardcore setting allows, opening up questions that are difficult, if not impossible, to formulate in the original hardcore language. In particular, in our bosonic model the form of nonlinear diffusion can be freely tuned, which enables us to study a version of the $A+B\rightarrow\emptyset$ process in which one species has a nonlinear bias while neither species has a linear bias, a setting in which the scaling behavior of the density had not previously been considered.

The flexibility of the bosonic setting further provides a convenient testing ground for exploring the relationship between the transport properties of the individual particle species in isolation and the scaling behavior of the total density in two-species pair-annihilation processes.
To this end, we also study a version of our bosonic model in which the biased nonlinear hopping occurs only through three-body reaction rules.
This variant still gives rise to $z=\frac32$ superdiffusion for each species in isolation, yet displays a $t^{-1/4}$ density decay instead.
This shows that KPZ scaling in the single-species model is not, in general, sufficient to produce $t^{-1/3}$ behavior in the corresponding pair-annihilation model, and we discuss a possible explanation for this observation based on scaling arguments from the Doi--Peliti field theory~\cite{D1976,D1976II,Gra1980,P1985}.

The rest of this paper is organized as follows. In Sec.~\ref{Sec:model}, we define the two-species bosonic pair-annihilation model and write down its field-theoretic action using the Doi-Peliti formalism.
In Sec.~\ref{Sec:KPZ}, we study the single-species case, which is related to the surface-growth problem. In particular, we discuss KPZ scaling from the point of view of the Doi-Peliti formalism and present numerical results confirming this expectation.
In Sec.~\ref{Sec:2bdy}, we study the pair-annihilation model, focusing in particular on the role of nonlinear diffusion, and show through numerical simulations that the hardcore effect observed in Refs.~\cite{Janowsky1995,Janowsky1995b,Is1995} can indeed be reproduced within a bosonic model.
In Sec.~\ref{Sec:3Bdy}, we perform a similar analysis on the model with purely three-body hopping and provide numerical evidence for KPZ scaling of the single-species problem alongside a $t^{-1/4}$ pair-annihilation density decay.
In Sec.~\ref{Sec:Conc}, we summarize our results and conclude the paper.

\section{\label{Sec:model}Model and Doi-Peliti formalism}
We consider a one-dimensional lattice system of size $L$ with periodic boundary conditions populated by two species of particles $A$ and $B$.  The number of  $A$ and $B$ particles at site $n$ will be be denoted by $\fA_n$ and $\fB_n$, respectively.  The occupation number of both species of particles at each site is unbounded. 

Both species of particles are allowed to hop to adjacent lattice sites with rates that depend non-linearly on the number of particles at the site they occupy.  This is achieved by the reaction rules, written in stoichiometric notation, $A_n\to A_{n\pm1}$ and $2A_n\to A_n+A_{n\pm 1}$, and similarly for $B_n$, by which a particle at site $n$ can hop to site $n\pm 1$ with the rates:
\begin{align}
\label{HoppingRule}
\begin{split}
R_{A}^\pm(\fA_n)=&D_{A}(1\pm \tfrac12 d_{A}) \fA_n \\
&+ Q_{A} (1\pm \tfrac12 q_{A}) \fA_n (\fA_n-1), \\
R_{B}^\pm(\fB_n)=&D_{B}(1\pm \tfrac12 d_{B}) \fB_n \\
&+ Q_{B} (1\pm \tfrac12 q_{B}) \fB_n (\fB_n-1).
\end{split}
\end{align}
In addition, particles of opposing species mutually annihilate according to the rule $A+B\to\emptyset$.  A pair-annihilation event at site $n$ occurs with rate $\sigma \fA_n \fB_n$, where 
$\sigma$ is a nonnegative constant.

The master equation describing this process can be expressed using the conventional bosonic formalism~\cite{Tauber2005}.
To this end, we note that a general configuration of particles may be encoded in the string of occupation numbers $\C=\fA_1,\dots,\fB_1,\dots$, which is naturally encoded in a bosonic Fock space as occupation basis vectors $|\C\rangle$. 
Within this framework, a probability distribution over states of the system is encoded in a general ket $|\P,t\rangle \coloneqq \sum_{\C} \P(\C,t) | \C\rangle$, where by $\P(\C,t)$ we denote the probability the system is in configuration $\C$ at time $t$.

The effect the reaction rules have on states is most conveniently expressed through the bosonic creation and annihilation operators,
\begin{align}\begin{split}
\hat a^\dag_n |\C\rangle &=  |\ldots, \fA_{n}+1,\ldots\rangle,\\
\hat b^\dag_n |\C\rangle &=  |\ldots, \fB_{n}+1,\ldots\rangle,\\
\hat a_n |\C\rangle &= \fA_n |\ldots, \fA_{n}-1,\ldots\rangle,\\
\hat b_n |\C\rangle &= \fB_n |\ldots, \fB_{n}-1,\ldots\rangle,
\end{split}\end{align}
which obey the canonical commutation relations $[\hat a_n,\hat a_m^\dagger]=\delta_{nm}=[\hat b_n,\hat b_m^\dagger]$ and $[\hat a_n,\hat b_m]=0=[\hat a_n^\dagger,\hat b_m]$.
Following the usual convention, the creation operators are denoted with daggers even though they are not the adjoints of the annihilation operators.

Using these, the master equation can be expressed through an equivalent imaginary-time Schr\"odinger equation,
\begin{equation}\label{Eq:master}
\partial_t|\P,t\rangle = - \hat H |\P,t\rangle
\end{equation}
where the effective (non-Hermitian) Hamiltonian can be written as a linear combination of terms corresponding to each local dynamic rule, $\hat H = \sum_n \hat H_n$, with $\hat H_n = \hat H_n^s + \hat H_n^p + \hat H_n^\sigma$,
\begin{align}\begin{split}
\hat H_n^s \coloneqq 
& -D_A \left ( \tilde \Delta \hat a_n^\dag +  d_A \tilde \partial \hat a_n^\dag
\right )\hat a_{n}\\&- D_B
 \left ( \tilde \Delta \hat b_n^\dag +  d_B \tilde \partial \hat b_n^\dag
\right )\hat b_{n},\\
\hat H_n^p \coloneqq 
& -Q_A\left ( \tilde \Delta \hat a_n^\dag +  q_A \tilde \partial \hat a_n^\dag
\right )\hat a_n^\dag \hat a_{n}^2\\&-
 Q_B\left ( \tilde \Delta \hat b_n^\dag + q_B \tilde \partial \hat b_n^\dag
\right )\hat b_n^\dag \hat b_{n}^2,\\
\hat H_n^\sigma \coloneqq &\sigma(- \hat a_n \hat b_n + \hat a_n^\dag \hat b_n^\dag \hat a_n \hat b_n),
\end{split}\end{align}
where $\tilde\Delta \mathcal{O}_n \coloneqq \mathcal{O}_{n+1} -2 \mathcal{O}_{n} +\mathcal{O}_{n-1}$ is the lattice Laplacian and $\tilde \partial \mathcal{O}_n \coloneqq (\mathcal{O}_{n+1} -\mathcal{O}_{n-1})/2$ is the lattice derivative.

Solutions of the master equation may be expressed as a path integral using the the Doi-Peliti formalism    .
To this end, we first introduce the left and right coherent states as
\begin{align}\begin{split}
    \langle a, b | &\coloneqq \sum_{\C} \langle \C | \prod_{n}(a_n^*)^{\fA_n} (b_n^*)^{\fB_n},\\
    |a,b\rangle &\coloneqq \sum_{\C} \prod_{n}\frac{a_n^{\fA_n} b_n^{\fB_n}}{\fA_n! \fB_n!} | \C \rangle,
\end{split}\end{align}
where $a_n$, $b_n$ are complex numbers and $z^*$ stands for the complex conjugate of 
$z$. Although $\langle a,b| \neq |a,b\rangle^\dag$, it is conventional to use this
notation. The resolution of unity is given by
\begin{align}
\int \left ( \prod_{n} \frac{d^2a_n d^2b_n}{\pi^2} e^{-|a_n|^2 - |b_n|^2}\right )
|a,b\rangle \langle a, b | = 1.
\label{Eq:unity}
\end{align}

With this, we seek a path-integral representation of the factorial-moment-generating function  $G(a_t,b_t)=\langle a_t,b_t|\P;t\rangle$ of the probability distribution $\P(\C,t)$.
For a purpose of succinct presentation, we use the subscript $t$ for the continuous 
variable $t$ for the collective parameters.
This satisfies:
\begin{equation}
\frac{\partial G}{\partial t} = - \mathcal{H}\left (a^*, b^*, \frac{\partial}{\partial a^*}, \frac{\partial}{\partial b^*} \right ) G,
\end{equation}
where $\mathcal{H}$ is obtained by replacing $\hat a^\dag$ and $\hat a$ by
$a^*$ and $\partial/\partial a^*$, respectively, (and similarly for $b$) in $\hat H$.

The solutions to this equation can be expressed by propagating an initial distribution 
$\P(\C,0)$ with the time evolution operator $\exp(-\hat H t) $, 
so that $G(a_t,b_t)=\langle a_t,b_t|e^{-\hat H t} |\P;0\rangle$. The latter may be brought into the form of a coherent state path integral through Trotterization by repeated insertion of Eq.~\eqref{Eq:unity}, resulting in:
\begin{equation}
\begin{split}
G(a_t,b_t) &= \int 
{\cal D}^2\{ab\}
e^{-S} \langle a_0, b_0| \P;0\rangle,
\\
S &= \sum_{n=1}^L \int_0^t dt' \left ( 
a_n^* \frac{\partial a_n}{\partial t'} +
b_n^* \frac{\partial b_n}{\partial t'} + H_n \right ),
\end{split}
\end{equation}
with $H_n$ to be obtained by replacing $\hat a^\dag$ by $a^*$ and $\hat a$ by $a$ (and similarly for $b$) in $\hat H_n$.
Taking the naive continuum limit gives the action $S = \int dt dx \cal L$, with the Lagrangian (density) given by:
\begin{align}
{\cal L} = &
a^*( \partial_t - D_A \partial_x^2 + v_A \partial_x )a 
+b^*( \partial_t - D_B \partial_x^2 + v_B \partial_x )b \nonumber \\
&-a^* a^2 (Q_A \partial_x^2 + u_A \partial_x )a^* 
-b^* b^2 (Q_B \partial_x^2 + u_B \partial_x )b^* \nonumber \\
&-\sigma (1- a^* b^*) a b,
\label{Eq:Ld}
\end{align}
where $a$, $b$ are fields in spacetime, $v_X = D_X d_X$ and $u_X = Q_X q_X$ for $X \in \{A, B\}$, and we work in units in which the lattice spacing is set to 1.

Scaling features of the model may be inferred from the continuum theory.
First, note that if $v_{A,B}=v$
then the ballistic terms $a^*\partial_x a$ and $b^* \partial_x b$ may be removed by a Galilean transformation $x\to x - vt$.
As a consequence, these terms do not affect the long-time and large-length behavior (although they can affect certain microscopic features).

Without this term, the Lagrangian coincides with that obtained in \cite{Lee1995} for linearly diffusing bosonic particles undergoing pair annihilation when $Q_{A,B}=0$.
There it was argued that the total density decays as $t^{-1/4}$ when starting from random initial conditions with the same number of both species.
In Sec.~\ref{Sec:2bdy}, we will numerically show that with a finite nonlinear bias this scaling is modified to $t^{-1/3}$.
The origin of this may be traced to the terms in Eq.\eqref{Eq:Ld} with the $u_X$ coupling.  In the Doi-shifted coordinates $a^* = 1 + \bar a$ and $b^* = 1 + \bar b$, this term becomes $u_A\bar a\partial_x a^2$ (and similarly for $B$). 
This has the same form as the convection term in the Burgers' equation, which in one space dimension is known to give rise to superdiffusive transport with $z=\frac32$ \cite{BurgersEq1,J1986,BurgersEq2}.
The modified decay exponent has the form $t^{-1/(2z)}$ and may thus be understood as a consequence of the superdiffusive fluctuations.
This was established for hardcore particles with biased hopping in \cite{Park2000,vW2001}, for which a similar convection term arises in the corresponding field theory description.
Numerical evidence supporting this fact for the present bosonic model will be provided in Sec.~\ref{Sec:AB0}.

\section{\label{Sec:KPZ}Single-species case and the corresponding surface growth}
Before we analyze the pair-annihilation model, we first study the case of a single species, which is achieved by setting $\sigma=0$ or $\fB_n=0$ for all $n$.
In isolation, the nonlinear hopping of the $A$ particles is an example of a zero-range process (ZRP), for which various exact results are known (for a review, see, e.g., Ref.~\cite{EH2005}). 
To use the known results, we rewrite the hopping rates as 
$R_A^\pm = R_A r_\pm$, where $R_A \coloneqq (R_A^+ + R_A^-)/2$ and $r_+ + r_- = 1$. 
For convenience, we introduce a random variable $\M := \sum_{n=1}^L \fA_n$ that counts the number of $A$ particles in configuration $\C$, and we always assume $\fB_n = 0$ for all $n$ in  this section. 
It has been proven (see Sec.~2.3 of
Ref.~\cite{EH2005}) that if $r_\pm$ are 
independent of the particle number, then the steady state distribution $P_s(\C)$  is
\begin{equation}
\label{Eq:Zs}
\begin{split}
P_s(\C) &= 
\delta_{N,\M}Z_{L,N}^{-1} \prod_{n=1}^L f(\fA_n),\\
f(\fA_n) &= \prod_{k=1}^{\fA_n} \frac{1}{R_A(k)} \text{ for } \fA_n \ge 1 \text{ and } f(0)=1,
\end{split}
\end{equation}
where
$\delta_{N,\M}$ is the Kronecker $\delta$ symbol, $N$ is the total number of particles in the system (which here is fixed by the initial condition), 
$Z_{L,N}$ is a normalization constant, and we have assumed $R_A(k) > 0$ for all positive integers $k$.

If $d_A = q_A$, then we have $r_\pm = \frac12 \pm \frac14 d_A$
and, therefore, we can use Eq.~\eqref{Eq:Zs} with
\begin{equation}
f(m) =\frac{1}{m!} \prod_{k=1}^{m} \frac{1}{D_A + Q_A (k-1)}.
\end{equation}
For example, if $Q_A = 0$ and $D_A>0$, then the steady state is a multinomial distribution
\begin{equation}
P_s(\C)= 
L^{-N}\frac{N!}{\fA_1!A_2!\cdots \fA_L!}\delta_{N,\M},
\end{equation}
as is easily anticipated.
If $D_A = Q_A$, then Eq.~\eqref{Eq:Zs} gives
\begin{equation}
P_s(\C)=
C_{L,N}^{-1} \left ( \frac{N!}{\fA_1!A_2!\cdots \fA_L!} \right )^2 \delta_{N,\M}, 
\end{equation}
with a normalization constant $C_{L,N}$. Except the case of $L=2$, no closed formula for $C_{L,N}$ is known, to the best of our knowledge.
Notice that hopping bias does not affect the steady state as long as $d_A = q_A$. Anticipating that Edwards-Wilkinson (EW) or KPZ scaling will be
determined by whether $q_A$ is zero or not, we note that this observation naturally
recovers the well-established fact that the steady states of the EW and KPZ
equations in one dimension should be identical.

Note that the case of $D_A=0$ with finite $Q_A$ gives rise to the special situation in which  $R_A(1)=0$ and thus hopping occurs only at sites with at least two particles. 
As a consequence, any configuration with 
$\fA_n \le 1$ for all $n$ is an absorbing state in which no motion is possible.
This limit can be mapped to a version of the two-species pair-annihilation model in Sec.~\ref{Sec:model}, following \cite{Jain2005}.
First, we redefine $\fA_n' = \fA_n - 1$, which gives 
\begin{equation}
R_{A'}^\pm =  
2 Q_A (1 \pm \tfrac12 q_A) \fA_n'
+ Q_A (1 \pm \tfrac12 q_A) \fA_n'(\fA_n'-1),
\end{equation}
and interpret a site with more than one $A$ particle ($\fA_n > 1$) as a site with $\fA_n'=\fA_n-1$ particles of species $A'$, and hopping of $A$ as hopping of $A'$.
In addition, we interpret an empty site as a site occupied by a single $B'$ particle, and a site occupied by a single $A$ particle as a site containing neither an 
$A'$ particle nor a $B'$ particle.
Within this mapping, the only meaningful initial conditions are those with at most one $B'$ particle at any site.
To enforce the constraint that $A'$ and $B'$ particles do not occupy the same site, 
we set $\sigma = \infty$, so that pair annihilation occurs instantaneously. 
Through this mapping we recover the original pair-annihilation model 
with $D_A = 2Q_A$, $d_A = q_A$, $D_B = Q_B = 0$, and $\sigma = \infty$.
When $N > L$, the steady-state distribution in terms of $A'$ particles is again given by Eq.~\eqref{Eq:Zs} with $D_A = 2Q_A$ and $N \mapsto N - L$. Notably, even though $D_A = 0$ in the original setting, an effective linear-diffusion term emerges when $N > L$. 

We now move back to the general case of $R_A(k)>0$ for $k \ge 1$.
Since the total number of particles is conserved, the one dimensional lattice allows for a mapping to a surface growth model by setting
\begin{equation}\label{HeightFieldMicro}
h_{n+1} - h_n = \fA_n - \rho_0,
\end{equation}
where $\rho_0 = \sum_{n=1}^L \fA_n /L$ is the density of particles for a realized initial configuration.
For simplicity, we assume that $\rho_0$ is a fixed number rather than a random variable, which fixes its value to the particle density $\rho_0=N/L$.
This definition may equivalently be expressed as $h_n - h_1 = \sum_{k=1}^{n-1} (\fA_k - \rho_0)$.
Note that the height configuration satisfies periodic boundary conditions, $h_{L+1} - h_1 = \sum_{n=1}^L (\fA_n - \rho_0) = 0$.

The scaling behavior of the process may be extracted from the long-time and large-length behavior of the the width, defined as
\begin{equation}
W^2(t,L)\coloneqq \frac{1}L \left \langle \sum_{n=1}^L (h_n-\bar h)^2 \right \rangle,\quad
\bar h \coloneqq \frac1L \sum_{n=1}^L h_n.
\end{equation}
Asymptotic scaling behavior of $W^2(t,L)$ is expected to be
\begin{equation}
\lim_{L\rightarrow\infty} W^2(t,L) \sim t^{2\beta},\quad
\lim_{t\rightarrow\infty} W^2(t,L) \sim L^{2\alpha}, 
\end{equation}
where $\alpha$ is the roughness exponent, $\beta$ is the growth exponent,
and $z \coloneqq \alpha/\beta$ is the dynamic exponent.
Since we can write $W^2(t,L) = \langle (h_1 - \bar h)^2 \rangle$
due to the translational invariance, 
we write the width in terms of $\fA_n$'s as
\begin{align}
W^2(t,L) = \left \langle \left [ \sum_{k=1}^L \left ( 1 - \frac{k}{L} \right ) 
\left ( \fA_k-\rho_0 \right ) \right ]^2 \right \rangle.
\end{align}
When the steady state can be written as in Eq.~\eqref{Eq:Zs}, 
we have
$\langle \fA_n \fA_m \rangle = \delta_{nm} \langle \fA_1^2 \rangle 
+ (1 - \delta_{nm} ) \langle \fA_1 \fA_2 \rangle$,
which allows the steady-state width to be written as:
\begin{align}
\label{Eq:ZRPst}
\frac{W^2(\infty,L)}{L-1} = \left \langle \fA_1^2 - \fA_1 \fA_2\right \rangle
\frac{2L-1}{6L} \\
+ \left \langle \fA_1 \fA_2 -\rho_0^2\right \rangle \frac{L-1}{4},
\nonumber
\end{align}
where averages are taken over the factorized distribution in Eq.~\eqref{Eq:Zs}.

We can get an exact formula for the case of $Q_A=0$. 
Using $\langle \fA_1^2 \rangle = \rho_0^2 + \rho_0 (1-1/L)$
and $\langle \fA_1 \fA_2 \rangle = \rho_0^2 - \rho_0 /L$ for the multinomial distribution,
we have
\begin{align}
\label{Eq:EWsatval}
W^2(\infty, L) = \rho_0 \frac{L}{12}\left (   1 - L^{-2} \right ),
\end{align}
and, accordingly, the expected $\alpha=\frac12$.

The dynamical scaling may be inferred from the continuum description.
In the absence of the $B$ species, the Lagrangian from Eq.\eqref{Eq:Ld} becomes:
\begin{equation}
{\cal L} = 
a^*(v \partial_x- D \partial_x^2 )a 
-a^* a^2 (Q \partial_x^2 + u \partial_x )a^* ,
\end{equation}
where for simplicity we drop the subscript $A$.
The conservation of the total particle number is encoded in the field theory as a $U(1)$ symmetry $a\to e^{i\theta}a$ and $a^* \to e^{-i\theta} a^*$; the corresponding Noether conserved charge is $a^* a$, which is equal to the local particle density $\fA$.

The long-time and large-distance behavior is controlled by the hydrodynamic mode governing the transport of the conserved particle density.
First, consider the saddle point equations:
\begin{equation}
\begin{split}
\frac{\delta S}{\delta a^*}=&(\partial_t+v\partial_x-D\partial_x^2)a-a^2(u\partial_x+Q\partial_x^2)a^*\\
&+(u\partial_x-Q\partial_x^2)(a^* a^2)=0,\\
\frac{\delta S}{\delta a}=&(\partial_t+v\partial_x+D\partial_x^2)a^*\\
&+2a^* a(u\partial_x+Q\partial_x^2)a^* =0.
\end{split}
\end{equation}
The physical saddle point is given by $a^* =1$ and $a=\rho_0$ (initial density).
This solution breaks the $U(1)$ symmetry, as do the entire manifold of additional ({\em unphysical}) saddle points obtained by phase rotations of this solution, $a^* =e^{-i\theta}$ and $a=e^{i\theta}\rho_0$.
As a consequence, fluctuations around the physical saddle point are described by a gapless mode which parametrizes local rotations in the saddle point manifold.
This is obtained by promoting the symmetry generator to a spacetime-dependent field $\theta\to i\bar\varphi(t,r)$.  Additionally allowing the density to deviate from its stationary value gives the new parametrization:
\begin{equation}
a(t,r)=e^{-\bar\varphi(t,r)}\phi(t,r),\quad a^*(t,r)=e^{\bar\varphi(t,r)},
\end{equation}
whose Jacobian is one.  Here, $\phi$ is the continuum field representing the local particle number $\fA_n$.
This parametrization, 
also known as the Grassberger transformation~\cite{Gra1982}, is the correct symmetry-based parametrization of the hydrodynamic density fluctuations.

This parametrization brings the action to the form:
\begin{multline}
S=\int dtdx\big(\bar\varphi(\partial_t+v\partial_x-D\partial_x^2)\phi-D\phi(\partial_x\bar\varphi)^2\\
+\bar\varphi(u\partial_x-Q\partial_x^2)\phi^2-Q\phi^2(\partial_x\bar\varphi)^2\big).
\end{multline}
Note that the first line, which corresponds to the linear hopping processes $A_n\leftrightarrow A_{n+1}$, is the Martin-Siggia-Rose-Janssen-de Dominicis (MSRJD) representation of the Dean-Kawasaki equation \cite{Dean,Kawasaki}, which is known to describe the hydrodynamics of non-interacting identical particles undergoing independent random walks.

\begin{figure}
\includegraphics[width=\linewidth]{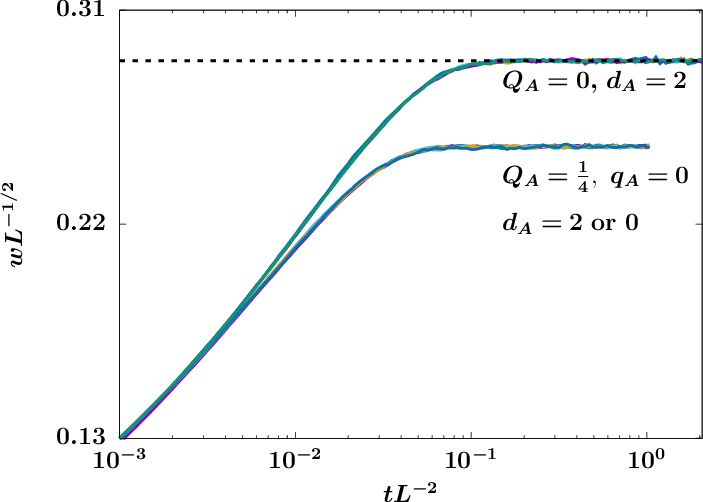}
\caption{\label{Fig:EW} Scaling-collapse plots of 
$w L^{-1/2}$ vs $tL^{-2}$ on a semi-logarithmic scale for
three sets of parameters ($Q_A=0$, $d_A=2$) (top curves), 
($Q_A=\frac14$, $q_A=0$, $d_A=2$), and 
($Q_A=\frac14$, $q_A=0$, $d_A=0$). $D_A$ is fixed $\frac12$ for all cases. The last two cases are graphically indistinguishable.
The system sizes are from $2^6$ to $2^{10}$, shown in different colors (though indistinguishable). 
The straight line indicates the value $1/\sqrt{12}$, which is obtained in Eq.~\eqref{Eq:EWsatval}.
}
\end{figure}
The scaling behavior of the process is obtained by expanding in small fluctuations around the stationary solution $\phi(t,r)=\rho_0+\varphi(t,r)$.  Passing to a moving frame $x\to x-(v+2\rho_0u)t$ and keeping only 
relevant terms in the sense of the renormalization group (RG) give:
\begin{equation}\label{S_Burgers}
S=\int dtdx\bigg(\bar\varphi(\partial_t-\tilde D\partial_x^2)\varphi-\frac{\nu}{2}(\partial_x\bar\varphi)^2
+\frac{\lambda}{2}\bar\varphi\partial_x\varphi^2\bigg),
\end{equation}
with $\tilde D=D+2\rho_0Q$, $\lambda=2u$, and $\nu=2D\rho_0+2Q\rho_0^2$.
This is the MSRJD action for the noisy Burgers' equation \cite{BurgersEq1,J1986,BurgersEq2}.
It is known to be dual to the KPZ equation via the mapping $\partial_xh=\varphi$, the continuum analogue of Eq.~\eqref{HeightFieldMicro}, and therefore leads to superdiffusive transport with the KPZ dynamical exponent $z=\frac32$.  The $\lambda=0$ limit, corresponding to tuning the rate of the non-linear hopping processes to zero, retrieves the MSRJD theory for EW diffusion.

Now we present numerical results, focusing on the Family-Vicsek scaling
\begin{equation}
w(L,t)\coloneqq\sqrt{W^2(L,t)} = t^\beta f(t/L^z),
\end{equation}
where $\beta = \frac14$, $z=2$ for the EW class and
$\beta = \frac13$ and $z=\frac32$ for the KPZ class.
To see how the width grows with time, we always use the initial condition that
$\fA_n=1$ for all $n$.
For simulating dynamics of bosons, we employ the method suggested in
Ref.~\cite{Park2005,Park2006}.

First, we consider the case of $q_A=0$. According to the anticipation of field theory, we expect to observe the EW scaling behavior regardless of whether $d_A$ is zero or not.
Indeed, as shown in Fig.~\ref{Fig:EW}, we have observed a nice scaling collapse with the
EW exponents $\alpha=\frac12$ and $z=2$. Hence, as the Galilean transformation argument
suggested, $d_A$ does not play any role in the scaling behavior.

\begin{figure}
\includegraphics[width=\linewidth]{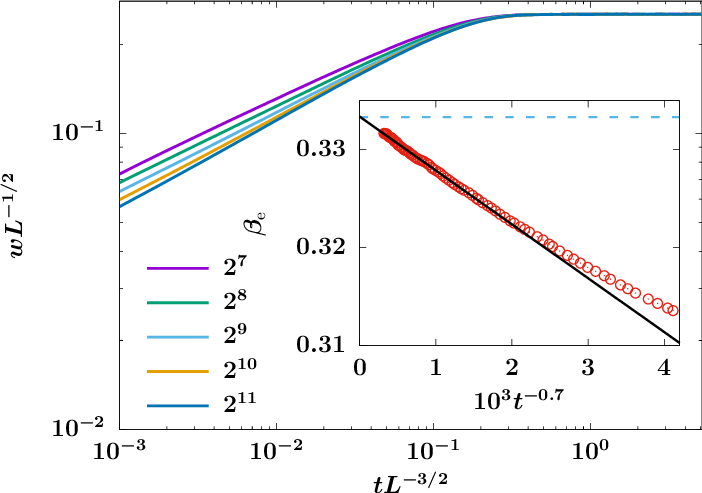}
\caption{\label{Fig:KPZ} Scaling-collapse plots of 
$w L^{-1/2}$ vs $t/L^{3/2}$ on a double logarithmic scale for
$Q_A=\frac14$, $q_A=2$, $D_A=\frac12$ for $L=2^7$ to $2^{11}$.
Inset: Plot of $\beta_e$ vs $t^{-0.7}$ with $b=32$. The system size is $L=2^{20}$, and the
maximum observation time is about $10^7$.
The dashed horizontal line indicates the value $\frac13$, and 
the solid straight line shows the result of a linear extrapolation.
Asymptotically, $\beta_e$ approaches $\frac13$, the growth exponent of the KPZ scaling.
}
\end{figure}
We expect the KPZ scailing when $q_A \neq 0$. To confirm this, we present simulation
results for $Q_A = \frac14$ and $q_A=2$ in Fig.~\ref{Fig:KPZ}.
Although the scaling collapse is not perfect for $L \le 2^{11}$, there is a clear
tendency that the collapse becomes better and better as $L$ increases.
In the inset of Fig.~\ref{Fig:KPZ}, we plot the effective exponent
\begin{align}
\beta_e(t) \coloneqq \log_b\frac{w(t)}{w(t/b)}
\end{align}
as a function of $t^{-0.7}$ for $L=2^{20}$, where  $b>1$ is a positive constant (not to be confused with the $B$ species Doi-Peliti field).  We drop the argument $L$, because finite-size effect up to the observation time is negligible.

The reason why $\beta_e$ is drawn as a function of $t^{-0.7}$ in the inset of Fig.~\ref{Fig:KPZ} is as follows.
Including corrections to scaling, the long-time behavior of $w(t)$ should be
\begin{align}
w(t) = C_0 t^{\beta} \left ( 1 + c t^{-\chi} + o(t^{-\chi} )\right ),
\end{align}
where $C_0$ and $c$ are non-universal constants whose precise numerical value is unimportant.
Hence, the long-time behavior of $\beta_e(t)$ should be
\begin{align}
\beta_e(t) = \beta - c \frac{b^{\chi}-1}{\ln b} t^{-\chi} + o(t^{-\chi}).
\label{Eq:chi}
\end{align}
Hence, if we plot $\beta_e$ as a function of $t^{-\chi} (b^\chi - 1)/\ln b$
with a correct $\chi$, the effective exponent in the asymptotic regime
becomes a linear function and does not depend on $b$, which will be used also in later numerical analysis.
Using the numerical method suggested in Ref.~\cite{P2013,P2014}, we get the information of $\chi$ by  analysizing
\begin{align}
\ln \frac{w(t) w(t/b^2)}{w(t/b)^2} = c (b^\chi-1)^2 t^{-\chi} + o(t^{-\chi}).
\end{align}
The numerical analysis of the above function for numerical result of $w(t)$ shows $t^{-0.7}$ behavior (details not shown here), which is the reason of choosing $t^{-0.7}$ as an abscisa of the inset of Fig.~\ref{Fig:KPZ}.
A linear extrapolation shows that
$\beta_e$ approaches $\frac13$, the growth exponent of the KPZ class, as expected.

We note that this behavior is not limited to specific model we consider, defined by Eq.~\eqref{HoppingRule}, but is rather a general feature of nonlinear hopping processes with a directional bias.
To illustrate this, it is worthwhile to discuss a recent study in \cite{Yama2025} within our framework.
In this work, the authors study a quantum Lindblad master equation defined with jump operators $\hat L_n=\sqrt\gamma\hat a_{n\pm1}^\dagger\hat a_n$ (where  $\hat a$ and $\hat a^\dagger$ in this context are the {\em quantum} bosonic creation and annihilation operators, distinct from those used elsewhere in this paper).
This model is exactly equivalent (via mapping at the operator level using the formalism in \cite{QuantumPopulations}) to the classical nonlinear hopping process with, using our terminology, $R_A^+ = \gamma \fA_n ( 1 + \fA_{n+1})$ and $R_A^-=0$,
which is described by the master equation with Hamiltonian
\begin{equation}
\hat H = - \gamma \sum_{n=1}^L \left ( \hat a_{n+1}^{\dag} -\hat a_n^\dag \right ) \hat a_n  \left ( 1 + \hat a_{n+1}^\dag \hat a_{n+1} \right ).
\end{equation}
In the stoichiometric notation, this is equivalent to the rules $A_n\to A_{n+1}$ and $A_n+A_{n+1}\to2A_{n+1}$.
In the literature, this type of process is  known as a misanthrope process 
(see Sec. 6.2 of Ref.~\cite{EH2005}).

From the theory about the misanthrope process in Ref.~\cite{EH2005},
we can conclude that the steady-state distribution $P_s(\C)$ becomes a constant
for any configuration. Therefore, if there are $N$ particles, then (recall that $\M$ is the number of
$A$ particles in configuration $\C$)
\begin{align}
    P_s(\C) = \binom{N+L-1}{N}^{-1} \delta_{N,\M},
\end{align}
which gives
\begin{equation}
\label{Eq:misan}
\langle\fA_1^2\rangle=\rho_0\frac{L(1+2\rho_0)-1}{L+1},\,\,\langle\fA_1\fA_2\rangle=\rho_0\frac{L\rho_0-1}{L+1},
\end{equation}
where $\rho_0 = N/L$.
Plugging Eq.~\eqref{Eq:misan} into Eq.~\eqref{Eq:ZRPst}, we get
\begin{equation}
W^2(\infty,L) = \frac{\rho_0(1+\rho_0)}{12} (L-1).
\end{equation}
If we set $\rho_0 = \frac12$, we get $w \sim \sqrt{L}/4$ which is
consistent with numerical results in Fig.~2(c) of Ref.~\cite{Yama2025}.

The same field theory treatment as discussed above can be applied to this set of rules; at finite density the result is indeed a noisy Burgers' equation, as claimed in Ref.~\cite{Yama2025}.  The corresponding MSRJD action, after passing to the moving frame $x\to x-\gamma(1+2\rho_0)t$, has the same form as Eq.~\eqref{S_Burgers} with 
$\tilde D=\gamma/2$, $\nu=\gamma\rho_0(1+\rho_0)$, and $\lambda=2\gamma$.

A similar analysis may be performed for any non-linear hopping process to which the Doi-Peliti formalism applies; see Sec.~\ref{Sec:3Bdy}.  When the non-linear processes lack spatial parity symmetry, the long-time and large-length effective action will generically be of the form of Eq.~\eqref{S_Burgers} and so the transport will generally be KPZ superdiffusive.
Spatially parity-symmetric processes will have $\lambda=0$, and thus will undergo EW diffusion.

An exception occurs when $R_A(k)=0$ for $k \le k_0$ and $R_A(k)\neq 0$ for $k>k_0$ for some positive integer $k_0$.  At low densities $\rho_0 < k_0$, this hopping rule gives rise to infinitely many absorbing states in which $\fA_n\le k_0$ for all $n$. In such cases, stationary expectations of large powers of $a_n$ vanish, $\langle a_n^{k_0+1}\rangle = 0$, while low powers such as $\langle a_n \rangle$ need not vanish. This discrepancy arises from the fact that $\langle a_n^{k_0+1} \rangle = \langle \fA_n (\fA_n - 1) \cdots (\fA_n - k_0)\rangle$, which significantly deviates from $\langle a_n \rangle^{k_0+1} = \langle \fA_n \rangle^{k_0+1}$ in one of the absorbing states. Consequently, the saddle-point equation around whose solution the perturbative expansion was performed fails to provide a proper basis for the perturbative field theory.  Since the Doi-Peliti formalism is exact (once the path integral is properly treated as a limit), this failure is due not to the formalism itself but to the lack of a physically meaningful solution of the saddle-point equation.

The ZRP with $D_A=0$ discussed above provides an example with $k_0=1$.  As we have already discussed, this model may be mapped to a model with two different species $A'$ and $B'$, and accordingly two independent fields $a_n'$ and $b_n'$, even though the original model is of a single species. This mapping extends straightforwardly to the case of $k_0>1$.
More generally, this scenario provides a class of examples in which conventional hydrodynamics breaks down, and instead an emergent dynamics involving multiple composite degrees of freedom becomes the correct long-distance description. 
This scenario is not unprecedented in models with an absorbing state. A notorious example is the pair contact process with diffusion; see Ref.~\cite{PP2005}.
On the other hand, there is no impediment to applying the above procedure when $\rho_0>k_0$.  In this scenario, the usual linear diffusion term $\bar\varphi\partial_x^2\varphi$ is generated (as we have already shown using the above mapping) by the expansion around the solution of the saddle-point equation and the hydrodynamic description emerges in the usual way.

\section{\label{Sec:AB0}Pair annihilation with nonlinear bias}
\subsection{\label{Sec:2bdy}Two-body hopping}
Now we return to the pair annihilation problem. As was shown in the previous section, the biased nonlinear hopping leads to the KPZ scaling of each species independently.  We now show numerically that this superdiffusive transport leads to a modification to the power law decay of the density.

The initial condition is chosen such that each species has the same initial population, which is taken to be independently multinomially distributed over the lattice, 
\begin{equation}
\label{Eq:ABinit}
\P(\C,t=0) = 
\frac{(N!)^2}{L^{2N}} \prod_{n=1}^L \frac{1}{\fA_n! \fB_n!},
\end{equation}
where only the case of $\sum_{n}\fA_n = \sum_n \fB_n=N$ is under consideration.

Our main observable is the density of total particles at time $t$, defined  as
\begin{align}
\rho(t) \coloneqq \frac1{L} \sum_{n=1}^L \left \langle \fA_n + \fB_n \right \rangle,
\end{align}
where $\langle \cdot \rangle$ stands for average over all realizations for a given initial condition.

With the initial condition Eq.~\eqref{Eq:ABinit}, we expect a power-law decay as
\begin{equation}
\rho(t) \sim t^{-\delta},
\end{equation}
for an infinite system.

\begin{figure}
\includegraphics[width=\linewidth]{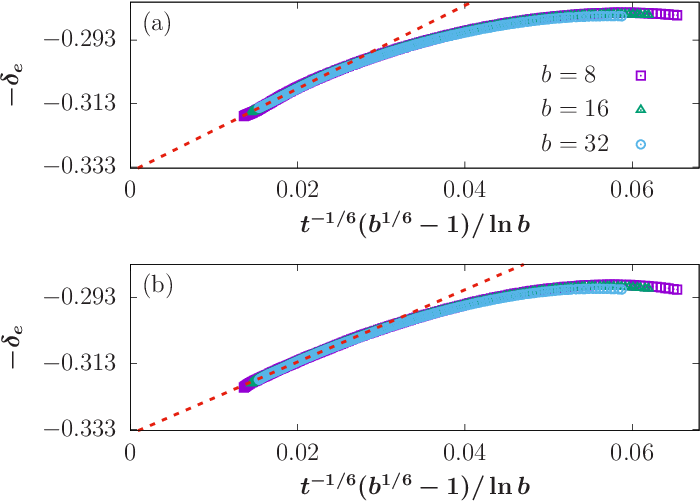}
\caption{\label{Fig:delta} Plots of $-\delta_e$ against 
$t^{-1/6}  ( b^{1/6} - 1 )/\ln b$ with $b=8$ (square), 16 (triangle), and 32 (circle)  for $D_A=D_B=\frac12$, $d_A=d_B=0$, $Q_A=\frac14$, $q_A=2$, and (a) $Q_B=\frac14$, $q_A=2$ or (b) $Q_B=0$.
The results of linear extrapolation in each panel are drawn as dashed lines, to give $\delta \approx \frac13$ for
both cases.}
\end{figure}

We now present simulation results for the density-decay exponent.
In the simulations presented below, $\sigma=1$ and $\rho(0)=2$ are always assumed.
To estimate $\delta$, we study the effective exponent 
$-\delta_e (t;b)$, defined as
\begin{align}
-\delta_e (t;b) := \log_b \frac{\rho(t)}{\rho(t/b)}.
\end{align}
We can repeat the same analysis as Eq.~\eqref{Eq:chi} for $-\delta_e$.
That is, if we depict $-\delta_e$ as a function of $t^{-\chi} (b^\chi-1)/\ln b$
with the correct $\chi$, all effective exponents for different $b$ should  collapse onto a single straight line for large $t$ (or small $t^{-\chi}$). 
In Fig.~\ref{Fig:delta}(a), we present the simulation results of $-\delta_e$ for $D_A = D_B=\frac12$, $d_A=d_B=0$, $Q_A = Q_B=\frac14$, $q_A = q_B= 2$.
Here, the system size is $L=2^{22}$, maximum observation time is $t \approx 10^7$, and the number of independent runs is about $100$.
We set $\chi = \frac16$ as suggested in Ref.~\cite{Is1995}. Indeed, the bosonic model with 
nonlinear bias reproduces $\delta=\frac13$.

In the numerical analysis, we have used $\chi=\frac16$, but in Ref.~\cite{benAv} reported are simulation results that are consistent with the claim $\chi = 1/12$ in Ref.~\cite{Janowsky1995b}. Since corrections to scaling are not likely to be universal, we believe that the discrepancy of $\chi$ in our numerical results from that in Ref.~\cite{benAv} need not be considered contradictory, as long as $\delta$ is consistent in both studies. In fact, if we use $\chi=1/12$ and naively extrapolate the effective exponent, we get $\delta \approx 0.36$ (details not shown here), which is inconsistent with all the previous theories. 

A benefit of studying the bosonic model over hardcore particles is that we can make only one species in a cluster perform Burgers' equation behavior without the introduction of a linear bias to either species.
This is achieved by setting $D_A = D_B=\frac12$, $d_A=d_B=0$, $Q_A = \frac14$, $q_A = 2$, and $Q_B=0$. 
Since $A$-rich regions are supposed to spread as $t^{2/3}$ and  $B$-rich regions as $t^{1/2}$, the dynamics of the $A$-rich regions should dominate, suggesting $\rho(t) \sim t^{-1/3}$ even in this case. 
In Fig.~\ref{Fig:delta}(b), we depict the effective exponent obtained by simulations for this 
case, which indeed supports the anticipation $\delta = \frac13$.

To confirm that same linear bias for two species indeed does not affect the asymptotic behavior, we also performed simulations for the case of 
$D_A=D_B=\frac12$, $d_A=d_B=q_A=2$, $Q_A=\frac14$, and $Q_B=0$, to get almost indistinguishable results from Fig.~\ref{Fig:delta}(b) (details not shown here).

\subsection{\label{Sec:3Bdy}Three-Body Hopping}
Now we can address the question of whether the $t^{-1/3}$ power law always emerges when the single particle model (without the annihilation event) is in the KPZ class.
To this end, we modify the hopping rates as
\begin{align}
    \label{Eq:ThreeKPZ}
R_A^+(\fA_n)=\gamma_1A_n + \frac{\gamma_3}{6} \fA_n (\fA_n-1)(\fA_n-2),
\end{align}
and $R_A^-=0$ for species $A$; and the same form of rates for species $B$.
In the stoichiometric notation, this is equivalent to having a three-body hopping rule $3A_n\to2A_n+A_{n+1}$ but no two-body hopping, in addition to the linear process $A_n\to A_{n+1}$.

\begin{figure}
    \includegraphics[width=\linewidth]{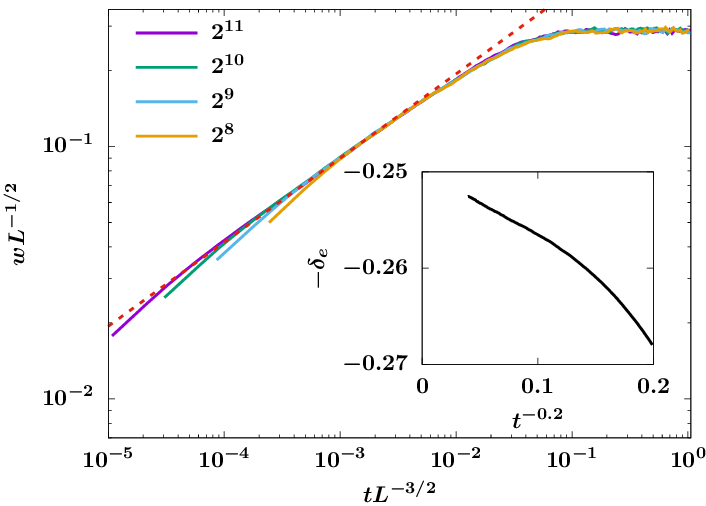}
\caption{\label{Fig:TC} Scaling collapse of $w L^{-1/2}$ vs $t/L^{-3/2}$ for the single
species model with hopping rate Eq.~\eqref{Eq:ThreeKPZ} for $L=2^8$, $2^9$, $2^{10}$, and $2^{11}$.
The collapse looks perfect, which supports the KPZ scaling. For comparison, we also draw
a straight line with slope $\frac13$ (red dashed line). Inset: Plot of $-\delta_e$ vs $t^{-0.2}$ for the pair annihilation model with hopping rate Eq.~\eqref{Eq:ThreeKPZ}. Here, $b=32$ is set. For both simulations, $\gamma_{1,3}=1$}
\end{figure}
If no $B$ particles are present, which becomes a ZRP, the Doi-Peliti field theory with changes of variables $a^* = e^{\bar \varphi}$ and $a = e^{-\bar\varphi} (\rho_0 + \varphi)$ predicts that this model should belong to the KPZ class.
Indeed, repeating the field theory analysis discussed above retrieves, in the moving frame $x\to x-(\gamma_1+3\rho_0^2\gamma_3)t$, Eq.~\eqref{S_Burgers} with the transport coefficients $\tilde D=(\gamma_1+3\gamma_3\rho_0^2)/2$, $\nu=\rho_0\gamma_1+\rho_0^3\gamma_3$, and $\lambda=6\rho_0\gamma_3$.
To confirm this anticipation, we simulate this model with the initial condition, $\fA_n=1$ and $\fB_n=0$ for all $n$.
The resulting scaling collapse plots are presented in Fig.~\ref{Fig:TC}, to observe indeed the KPZ scaling behavior.

Now we move on to the case of nonzero $\sigma$.
As before, we use the initial condition Eq.~\eqref{Eq:ABinit} with $\rho(0)=2$.
In simulations, we set $L=2^{22}$ and time evolves up to $10^7$. 
The average density is obtained over about 100 independent runs. The effective exponent $-\delta_e$ is depicted in the inset of Fig.~\ref{Fig:TC}. As the inset of Fig.~\ref{Fig:TC} shows, the density decays as $t^{-1/4}$ even though the density-conserved version belongs to the KPZ class.

The difference between this and the two-body hopping process considered previously may be explained by examining the difference between the Doi-Peliti action in Eq.~\eqref{Eq:Ld} and that corresponding to Eq.~\eqref{Eq:ThreeKPZ}.
In the Doi-shifted coordinates, the nonlinear hopping in Eq.~\eqref{Eq:ThreeKPZ} contributes a cubic advection term $\bar a \partial_x a^3$.
Unlike the single species case at finite density, there is no background density to expand around because the stationary point of the pair annihilation process is $a=0$.  Near the pair annihilation fixed point, the fields $a$ and $\bar a$ respectively have scaling dimension $-d=-1$ in one dimension and $0$.  Consequently, the cubic advection term is irrelevant in the RG sense. This contrasts with the the Burgers' advection term $\bar a\partial_xa^2$ that comes from two-body hopping term, which is marginal in one dimension and leads to modified scaling.
This implies that in the absence of two-body hopping terms the scaling behavior is determined entirely by the linear diffusive fluctuations, resulting in the density decay $\rho \sim t^{-1/4}$ rather than $t^{-1/3}$.

\section{Conclusion}\label{Sec:Conc}
We have numerically demonstrated that $A+B\to\emptyset$ pair annihilation of bosonic particles whose motion is determined by certain ZRPs exhibit a power law $t^{-1/3}$ decay of the total density.
We thus confirm that this behavior is not unique to particles with hardcore exclusion, despite what was previously argued \cite{Janowsky1995,Is1995}.
We found the essential ingredient necessary to produce this modified scaling in bosonic models to be a nonlinear hopping processes with a directional bias, so that the corresponding Doi-Peliti field theory contains a Burgers' advection term $\bar a\partial_xa^2$.

We also showed that while directionally-biased ZRPs with only three-body nonlinear lead to superdiffusive transport in the KPZ universality class with $z=\frac32$, two-species $A+B\to\emptyset$ pair annihilation of particles undergoing this type of motion display a total density which decays like $t^{-1/4}$, which is associated with diffusive fluctuations.
This demonstrates that the transport features of particle species in isolation do not alone determine the density-decay exponent in the $A+B\to\emptyset$ processes.
A more comprehensive study of how anomalous transport of individual species affects pair-annihilation dynamics is left to future work.

\begin{acknowledgments}
We thank Joachim Krug and Sebastian Diehl for helpful discussions. S.-C.P. acknowledges support by the National Research Foundation of Korea (NRF) grant funded by the Korea government (MSIT) (No. RS-2026-25493117); and by the Research Fund of The Catholic University of Korea in 2026. 
F.T. is supported by the Deutsche Forschungsgemeinschaft (DFG, German Research Foundation) under Germany's Excellence Strategy Cluster of Excellence Matter and Light for Quantum Computing (ML4Q) EXC 2004/1 390534769 and by the DFG Collaborative Research Center (CRC) 183 Project No. 277101999 - project B02. We are grateful to the Center for Advanced Computation at KIAS for help with computing resources.
\end{acknowledgments}

\bibliography{abs.bib}

\end{document}